\documentclass[12pt]{iopart}
\usepackage{iopams,psfig,amssymb}
\eqnobysec

\newcommand{\figwidth}{0.9\columnwidth}

\begin{document}
\jl{1}

\title{Integrable impurities for an open fermion chain}

\author{Xi-Wen Guan\dag\ddag,
        Uwe Grimm\dag, Rudolf A R\"omer\dag\ and Michael Schreiber\dag}

\address{\dag\ Institut f\"ur Physik, Technische
  Universit\"at, D-09107 Chemnitz, Germany}

\address{\ddag\ Department of Physics, Jilin University,
         Changchun 130023, China}

\begin{abstract}
  Employing the graded versions of the Yang-Baxter equation and the
  reflection equations, we construct two kinds of integrable
  impurities for a small-polaron model with general open boundary
  conditions: (a) we shift the spectral parameter of the local Lax
  operator at arbitrary sites in the bulk, and (b) we embed the
  impurity fermion vertex at each boundary of the chain. The
  Hamiltonians with different types of impurity terms are given
  explicitly. The Bethe ansatz equations, as well as the eigenvalues
  of the Hamiltonians, are constructed by means of the quantum inverse
  scattering method. In addition, we discuss the ground-state
  properties in the thermodynamic limit. 
\end{abstract}


\submitted

\maketitle

\section{Introduction}
\label{sec1}

The interplay between disorder and many-body interactions continues to
be one of the main topics of condensed matter physics. In
one-dimensional quantum many-body systems without disorder, the Bethe
ansatz (BA) has proven to be a valuable tool, giving access to the energy
spectra and eigenstates of certain so-called integrable models
\cite{Mattis,EK}. At first glance, it appears that the integrability,
namely the existence of infinitely many integrals of motion, seems to
preclude any applications of the method to disordered
systems. However, in 1984 it was shown \cite{Andrei,Affleck} how to
apply the BA method to the Kondo problem \cite{Kondo} of a
single magnetic impurity in a bath of conduction electrons. Further
developments led to the construction and solution of integrable spin
chains with embedded spin defects \cite{spindefect,AJ}.

A different approach to integrable impurity models was considered in
references \cite{shift} where impurity vertices are
introduced by varying the {\em local}\/ interaction parameters while
preserving integrability. These studies have stimulated further
investigations \cite{Schmitteckert,EPR} of such impurities in various
systems. The resulting models have impurity terms with couple to the
charge degrees of freedom, and look fairly similar to generic impurity
terms. However, the energy spectrum is independent of the spatial
distribution of the defects, and there is no localization of the
ground-state wave function, unlike what is expected for generic
impurities. This peculiar behaviour can be understood by the fact that
integrability implies a purely forward-scattering mechanism at the
impurities \cite{EPR}. There is no reflection and thus no possibility
of destructive quantum interference which could lead to a localization.

Back-scattering can be introduced into integrable models by choosing
suitable boundary conditions (BC). Sklyanin \cite{Sklyanin} proposed a
systematic approach to construct and solve integrable quantum spin
systems with {\em open} BC. Central to his method are the so-called
reflection equations (RE) \cite{bound} which are the boundary
analogues of the Yang-Baxter equations (YBE) \cite{YBE}. Together, the
YBE and RE imply the integrability of a model which can then be
constructed as usual by the algebraic approach of the quantum inverse
scattering method (QISM) \cite{QISM}. The finite-size corrections of
the corresponding energy spectra and the asymptotic behaviour of
correlation functions follow predictions based on boundary conformal
field theory \cite{BCFT}. We remark that the BC of Sklyanin
\cite{Sklyanin} are called ``open'' in order to distinguish them from
the more often used periodic and the free BC. Although the term
``open'' seems to suggest particular transmission and reflection
properties, this is not necessarily implied. The combination of open
BC and integrable impurities has been considered in references
\cite{ref1,ref2,ref2w,bi}. Of particular interest is the case where the
forward-scattering impurity is directly coupled to a back-scattering
open boundary \cite{ref2w,bi}.  This combination may lead to physically
relevant, yet completely integrable models.

In the present article, we construct two kinds of integrable
impurities for a fermionic small-polaron model with general open BC.
Due to the fermionic nature of the model, we employ the graded version
of the QISM \cite{gradQISM,ref11}. For well-separated impurity
vertices located within the bulk, the local interaction terms involve
the two neighbouring sites as usual \cite{shift}. Placing the
forward-scattering impurities at the back-scattering boundaries, we
derive a Hamiltonian with rather general boundary terms which may be
interpreted as sources and sinks of particles at the boundaries.
Using the graded YBE and the graded RE, we derive the BA equations,
and obtain expressions for the eigenvalues for special cases of the
Hamiltonian. In addition, we discuss the ground-state properties in
the thermodynamic limit.

The paper is organized as follows. In section~\ref{sec2}, we introduce
the small-polaron model with general open BC. In section~\ref{sec3}, a
class of integrable impurities is constructed by shifting the spectral
parameters of local Lax operators at arbitrary sites in the bulk.  By
embedding the impurity fermion vertex at each boundary of the model,
we construct a class of integrable impurities with perfect
back-scattering in section~\ref{sec4}.  In section~\ref{sec5}, we
study the algebraic BA solutions for those impurity models.  The
ground-state properties are discussed in section~\ref{sec6}.
Section~\ref{sec7} is devoted to a discussion and conclusion.

\section{The small-polaron model}
\label{sec2}

We consider the small-polaron model \cite{ref6}, which describes the
motion of an additional electron in a polar crystal. The Hamiltonian
reads
\begin{eqnarray}
\bi{H} & = & -
J\sum_{j=2}^{N}(\bi{a}_j^{\dagger}\bi{a}_{j-1}^{}+
                \bi{a}_{j-1}^{\dagger}\bi{a}_j^{})\; +\;
V\sum_{j=2}^{N}\bi{n}_{j}^{}\bi{n}_{j-1}^{} \; +\;
W\sum_{j=1}^{N}\bi{n}_{j}^{} \nonumber\\ & &
\mbox{}+\; p_{+}\bi{n}_{N}^{}\;  +\; \alpha_{+} \bi{a}_{N}^{\dagger}\; +\;
\beta_{+} \bi{a}_{N}^{} \; +\; p_{-}\bi{n}_{1}^{}\;+\;
\alpha_{-}\bi{a}_{1}^{\dagger}\;+\;\beta_{-}\bi{a}_{1}^{},
\label{2a}
\end{eqnarray}
where $J$ is proportional to the overlap integral, $V$ denotes
the electron-phonon coupling and $W$ is the energy of the polaron.
The boundary coefficients $p_{\pm}$, $\alpha_{\pm}$ and $\beta_{\pm}$ are
Grassmann variables, with $p_{\pm}$ even and $\alpha_{\pm}$,
$\beta_{\pm}$ odd. Hermiticity of the Hamiltonian requires
$\alpha_{\pm}^{\dagger}=\beta_{\pm}$ and $p_{\pm}^{\dagger}=p_{\pm}$.
The fermionic creation and annihilation operators $\bi{a}_{j}^{\dagger}$
and $\bi{a}_{j}^{}$ satisfy the usual anticommutation relations
\begin{equation}
\{ \bi{a}_{j}^{},\bi{a}_{k}^{}\} \; = \;
\{ \bi{a}^{\dagger}_{j},\bi{a}^{\dagger}_{k}\}\; =\; 0, \qquad
\{\bi{a}_{j}^{},\bi{a}^{\dagger}_{k}\} \; = \; \delta_{j,k},
\end{equation}
and $\bi{n}_j^{}=\bi{a}_{j}^{\dagger}\bi{a}_{j}^{}$.
The $R$-matrix and local monodromy matrix are explicitly given as \cite{ref7}
\begin{equation}
R_{12}(u_1,u_2)=\left(\matrix{{a}^{\prime\prime}_+&0&0&0\cr
0&-i{b}^{\prime\prime}_-&{c}^{\prime\prime}&0\cr 0&{c}^{\prime\prime}&i{b}^{\prime\prime}_+&0\cr
0&0&0&-{a}^{\prime\prime}_-\cr }\right)\label{R}
\end{equation}
and
\begin{equation}
\bi{L}_j(u) = \left(
  \matrix{{b}^{\prime}_{+}+(i{a}^{\prime}_{+}-b_{+}^{\prime})\bi{n}_{j}^{} &
          {c}^{\prime}\bi{a}_{j}^{}\cr
           -i{c}^{\prime}\bi{a}^{\dagger}_{j} &
           {a}^{\prime}_{-}-({a}^{\prime}_{-}+i{b}^{\prime}_{-})
            \bi{n}_{j}^{}\cr }\right),\label{M}
\end{equation}
respectively. They satisfy the graded Yang-Baxter algebra (YBA)
\begin{equation}
R_{12}(u_1,u_2)\stackrel{1}{\bi{T}}\!(u_1)\stackrel{2}{\bi{T}}\!(u_2) \; =\;
\stackrel{2}{\bi{T}}\!(u_2)\stackrel{1}{\bi{T}}\!(u_1)R_{12}(u_1,u_2),
\label{2b}
\end{equation}
where
\begin{equation}
\bi{T}(u)=\bi{L}_{N}(u)\cdots \bi{L}_{2}(u)\bi{L}_{1}(u),
\label{2c}
\end{equation}
and
\begin{equation}
\stackrel{1}{X} \equiv X\otimes_{S}\mbox{id}_{V_2};\quad
\stackrel{2}{X} \equiv \mbox{id}_{V_1}\otimes_{S}X.
\end{equation}
Here, $\otimes_{S}$ is the supertensor product
\begin{equation}
[A\otimes_{S}B]_{\alpha\beta ,\gamma\delta}=(-1)^{[P(\alpha)+P(\gamma)]
P(\beta)}A_{\alpha\gamma}B_{\beta\delta}
\end{equation}
with the parity $P(1)=0$, $P(2)=1$ such that the $R$-matrix correspond
to the ``null'' parity case $P(\alpha)+P(\beta)+P(\gamma)+P(\delta)=0$
\cite{gradQISM}. We parametrize the coupling parameters $J$, $V$ and
$W$ as
\numparts%
\begin{eqnarray}
J & = & 1, \\
V & = & -2c_{2}(0),\\
W & = & 2s_{2}(0)\tan(w)+2c_{2}(0).
\end{eqnarray}
\endnumparts%
The entries of the $R$-matrix \eref{R} and the monodromy matrix
\eref{M} are
\numparts%
\begin{eqnarray}
a^{\prime\prime}_{\pm} & = &
  \xi_{+}^{\pm 1}(u_1)\xi_{+}^{\mp 1}(u_2)s_{2}(u_1-u_2),\\
b^{\prime\prime}_{\pm} & = &
  \xi_{+}^{\pm 1}(u_1)\xi_{+}^{\pm 1}(u_2)s_{0}(u_1-u_2),\\
c^{\prime\prime}       & = & c^{\prime} \; = \; s_{2}(0),\\
a^{\prime}_{\pm}       & = & \xi_{+}^{\pm 1}(u)s_{2}(u),\\
b^{\prime}_{\pm}       & = & \xi_{+}^{\mp 1}(u)s_{0}(u),\\
\xi_{\pm}(u)           & = & \frac{c_{0}(u\pm w)}{c_{0}(u)c_{0}(w)},
\end{eqnarray}
\endnumparts%
where we introduced the convenient notation
\begin{equation}
s_{n}(u) \equiv \sin(u+n\eta), \quad
c_{n}(u) \equiv \cos(u+n\eta).
\end{equation}
Throughout the paper, we therefore use $\eta$ and $w$ 
for the parametrization of the model parameters $V$, and $W$.

In a previous article \cite{ref10}, we proved that the model
\eref{2a} is integrable under the conditions that the boundary $K_{\pm}$
supermatrices
\begin{equation}
K_{\pm}(u)=\left(\matrix{K_{11}^{\pm}&K_{12}^{\pm}\cr
                         K_{21}^{\pm}&K_{22}^{\pm}}\right)
\label{2d}
\end{equation}
take the form
\numparts%
\begin{eqnarray}
\fl K_{11}^{-}& = &
  \xi_{+}s_{0}(u-\psi_{-})
   \left[\xi^{2}_{-}s_{2}(u)-\xi^{2}_{+}s_{-2}(u)\right],\label{eq:ab1}\\
\fl K_{22}^{-}& = &
  \xi_{-}s_{0}(u+\psi_{-})
  \left[\xi^{2}_{-}s_{-2}(u)-\xi^{2}_{+}s_{2}(u)\right],\\
\fl K_{12}^{-} & = &
  -\frac{\alpha_{-}s_{0}(\psi_{-})s_{0}(u)}{i\xi_{+}\xi_{-}s^{2}_{2}(0)}
  \left[\xi^{2}_{-}s_{2}(u)-\xi^{2}_{+}s_{-2}(u)\right]
  \left[\xi^{2}_{+}s_{2}(u)-\xi^{2}_{-}s_{-2}(u)\right],\\
\fl K_{21}^{-} & = &
  -\frac{\beta_{-}s_{0}(\psi_{-})s_{0}(u)}{is_{2}^{2}(0)}
  \left[\xi^{2}_{-}s_{2}(u)-\xi^{2}_{+}s_{-2}(u)\right]
  \left[\xi^{2}_{+}s_{2}(u)-\xi^{2}_{-}s_{-2}(u)\right],\\
\fl K_{11}^{+} & = &
  \xi_{+}s_{2}(u-\psi_{+}) \left[\xi^{2}_{-}s_{4}(u)-
                                 \xi^{2}_{+}s_{0}(u)\right],\\
\fl K_{22}^{+}& = &
\xi_{-}s_{2}(u+\psi_{+}) \left[\xi^{2}_{+}s_{4}(u)-
                                     \xi^{2}_{-}s_{0}(u)\right],\\
\fl K_{12}^{+} & = &
  -\frac{\alpha_{+}s_{0}(\psi_{+})s_{2}(u)}{i\xi_{+}\xi_{-}s^{2}_{2}(0)}
  \left[\xi^{2}_{-}s_{4}(u)-\xi^{2}_{+}s_{0}(u)\right]
  \left[\xi^{2}_{+}s_{4}(u)-\xi^{2}_{-}s_{0}(u)\right],\\
\fl K_{21}^{+} & = &
  -\frac{\beta_{+}s_{0}(\psi_{+})s_{2}(u)}{is^{2}_{2}(0)}
  \left[\xi^{2}_{-}s_{4}(u)-\xi^2_{+}s_{0}(u)\right]
  \left[\xi^{2}_{+}s_{4}(u)-\xi^2_{-}s_{0}(u)\right].
\label{eq:ab2}
\end{eqnarray}
\endnumparts%
Here we would like to emphasize that although we \cite{ref10}
construct the general boundary $K$-matrices \eref{2d} for the
small-polaron model \eref{2a} by the Lax pair formulation, we did not
figure out the form of the RE corresponding to more general boundary
$K$-matrices \eref{2d}.  In the above expressions, we further defined
\begin{equation}
p_{\pm}=s_{2}(0)\cot\psi_{\pm}. \label{para}
\end{equation}
The parameters $\psi_{\pm }$ control the strength of the boundary
potential terms, whereas $\alpha_{\pm}$ and $\beta_{\pm }$ in
\eref{eq:ab1}--\eref{eq:ab2} are the parameters characterizing the
fermion sources and sinks at the boundaries.  The Hamiltonian
\eref{2a} can be obtained as usual as an invariant of the commuting
family of transfer matrices $\btau(u)$
\begin{equation}
\btau (u)={\rm Str}_0[K_+(u)\bi{T}(u)K_-(u)\bi{T}^{-1}(-u)]\label{2e}
\end{equation}
 by taking the derivative at a special value of the spectral parameter
$u$. Namely,
\begin{equation}
-s_{2}(0)\:\left. \frac{d}{du}\btau(u)\right|_{u=0} = 2 \bi{H} \btau(0) +
{\rm Str}_{0}\left(\left.\frac{d}{du}K_{+}(u)\right|_{u=0} \right) ,
\label{relat}
\end{equation}
with ${\rm Str}_{0}$ denoting the supertrace with respect to the
auxiliary space.

\section{Integrable impurities in the bulk }
\label{sec3}

\begin{figure}
\centerline{\psfig{file=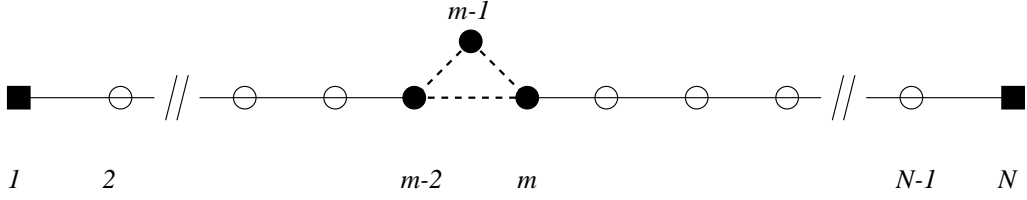,width=\figwidth}}
\caption{\label{fig-bmi} Graphical representation of the
  interactions in the chain with the impurity located at an arbitrary
  site $m$ together with boundary impurities.}
\end{figure}

In this section, we construct integrable impurities which appear in
the bulk part for the fermionic small-polaron model with general open BC.
If the quantum $R$-matrix of a fermionic system has the difference
property of spectral parameters, the associated Lax operator with
additional parameter also satisfies the graded YBA, i.e.,
\begin{equation}
  R_{12}(u_1-u_2)\stackrel{1}{\bi{L}}\!(u_1+\nu
  )\stackrel{2}{\bi{L}}\!(u_2+\nu ) \; =\;
  \stackrel{2}{\bi{L}}\!(u_2+\nu )\stackrel{1}{\bi{L}}\!(u_1+\nu
  )R_{12}(u_1-u_2).
\end{equation}
Therefore one can construct a class of integrable impurities for the
fermion model with both open and periodic BC by shifting the spectral
parameters of local monodromy matrices at arbitrary sites in the bulk.
The associated monodromy matrix is given as
\numparts%
\begin{eqnarray}
\bi{T}(u)
& =
& \bi{L}_{N}(u)\cdots \bi{L}_{m}(u+\nu_m)\cdots \bi{L}_{1}(u) \\
\bi{T}^{-1}(-u)
& =
& \bi{L}^{-1}_{1}(-u)\cdots \bi{L}^{-1}_{m}(-u+\nu_m)\cdots
\bi{L}^{-1}_{N}(-u),
\end{eqnarray}
\endnumparts%
where the parameter $\nu_{m}$ characterizes the impurity strength at
site $m$.  Now we suppose that the supermatrices $K_{\pm}$ are the
solutions of the graded RE \cite{ref4,gfy},
\numparts%
\begin{eqnarray}
\fl
{R_{12}(u_1-u_2) \stackrel{1}{K}_{-}\!(u_1) R_{21}(u_1+u_2)
 \stackrel{2}{K}_{-}\!(u_2) = } \nonumber \\
 \stackrel{2}{K}_{-}(u_2)
 R_{12}(u_1+u_2) \stackrel{1}{K}_{-}\!(u_1) R_{21}(u_1-u_2),
\label{3a} \\
\fl{
 R_{21}^{{\rm St}_1 {\rm iSt}_2}(-u_1+u_2)
 \stackrel {1}{K}\mbox{$\!$}_{+}^{{\rm St}_{1}}(u_1)
 R_{12}(-u_1-u_2-4\eta )
 \stackrel{2}{K}\mbox{$\!$}_{+}^{{\rm iSt}_{2}}(u_2) = } & & \nonumber \\
 \stackrel{2}{K}\mbox{$\!$}_{+}^{{\rm iSt}_{2}}(u_2)
 R_{21}(-u_1-u_2-4\eta )
 \stackrel{1}{K}\mbox{$\!$}_{+}^{{\rm St}_{1}}(u_1)
 R_{12}^{{\rm St}_1 {\rm iSt}_2}(-u_1+u_2),
\label{3b}
\end{eqnarray}
\endnumparts%
where $R_{21}(u)={\cal P}R_{12}(u){\cal P}$ with  the
graded permutation
\begin{equation}
{\cal P}_{\alpha \beta , \gamma \delta }=(-1)^{P(\alpha )P(\beta
  )}\delta_{\alpha \delta }\delta_{\beta \gamma },
\end{equation}
and ${\rm iSt}_{\alpha }$ denotes the inverse operation of the
supertransposition ${\rm St}_{\alpha }$ in the space $\alpha $.  The
$R$-matrix possesses the unitary property $R_{12}(u)R_{21}(-u)=\rho(u)
I,$ with $\rho (u)$ a scalar function of $u$.  It follows that the
double-row transfer matrix \eref{2e} may be considered as a generating
function of the infinite conserved quantities such that the
$K$-matrices are the solutions to the RE \eref{3a} and \eref{3b}.
{}From the relation \eref{relat}, it is not difficult to obtain the
Hamiltonian for the open fermion chain with an impurity located at
site $m$,
\begin{eqnarray}
\bi{H} & = &
\sum_{\stackrel{\scriptstyle j=2}{\scriptstyle j\ne m,m-1}}^{N}
 \bi{H}_{j,j-1} +
 \frac{1}{2}\bi{L}_{1}(0)K_{-}^{\prime}(0)\bi{L}_{1}^{-1}(0) +
 \frac{{\rm Str}_{0}[K_+(0)\bi{L}_{N}^{\prime}(0)\bi{L}_{N}^{-1}(0)]}
      {{\rm Str}_{0}[K_+(0)]} \nonumber\\
 & & +\bi{C}_{m,m-1}\bi{B}_{m,m-1}
     +\bi{A}_{m,m-1}\bi{H}_{m-1,m-2}\bi{B}_{m,m-1} \label{3d}
\end{eqnarray}
where
\numparts%
\begin{eqnarray}
\bi{H}_{j,j-1} & = &
  \bi{L}_{j-1}(0)\,\bi{L}_{j}^{\prime}(0)\,\bi{L}^{-1}_{j}(0)\,
  \bi{L}^{-1}_{j-1}(0),\\
\bi{A}_{m,m-1} & = &
  \bi{L}_{m-1}(0)\,\bi{L}_{m}(\nu_m)\,\bi{L}^{-1}_{m}(0)\,
   \bi{L}^{-1}_{m-1}(0), \\
\bi{B}_{m,m-1} & = &
   \bi{L}_{m-1}(0)\,\bi{L}_{m}(0)\,\bi{L}^{-1}_{m}(\nu_m)\,
   \bi{L}^{-1}_{m-1}(0), \\
\bi{C}_{m,m-1} & = &
   \bi{L}_{m-1}(0)\,\bi{L}_{m}^{\prime}(\nu_m)\,\bi{L}^{-1}_{m}(0)\,
   \bi{L}^{-1}_{m-1}(0).
\end{eqnarray}
\endnumparts%
The prime denotes the derivative with respect to the spectral
parameter $u$. The interactions of the open fermion chain with the
impurity are shown schematically in \fref{fig-bmi}.

In order to simplify the algebraic calculation for the construction of
such an integrable impurity for the model \eref{2a}, we let $w =0$.
The supermatrices \eref{2d} become (up to a normalization)
\numparts%
\begin{eqnarray}
K_-(u) = & \frac{1}{s_{0}(\psi_{-})} & \left(
              \matrix{-s_{0}(u-\psi_{-}) &  \alpha_{-}s_{0}(2u)\cr
                      \beta_{-}s_{0}(2u) & s_{0}(u+\psi_{-})\cr}
             \right), \label{3e}\\
K_+(u) = & & \left(
             \matrix{s_{2}(u-\psi_{+})  & \alpha_{+}s_{4}(2u)\cr
                     \beta_{+}s_{4}(2u) & s_{2}(u+\psi_{+})\cr }
             \right). \label{3f}
\end{eqnarray}
\endnumparts%
Then we check that the supermatrices \eref{3e} and \eref{3f} satisfy
the graded RE \eref{3a} and \eref{3b}, respectively.  From
\eref{3d}, after some algebra, we obtain the Hamiltonian for the
small-polaron model with both general open BC and an integrable impurity
located at site $m$ as
\begin{equation}
\fl
\bi{H} = \sum_{j=2}^{N} \bi{H}_{j,j-1}
+ \bi{H}_{1}^{\rm (b)} + \bi{H}_{N}^{\rm (b)}
+ \bi{H}_{m,m-1,m-2}^{\rm (h)} + \bi{H}_{m,m-1,m-2}^{\rm (d)}
+ \bi{H}_{m,m-1,m-2}^{\rm (c)} \label{3g}
\end{equation}
with
\numparts%
\begin{eqnarray}
\fl \bi{H}_{j,j-1} & = &
  \bi{a}_j^{\dagger}\bi{a}_{j-1}^{} + \bi{a}_{j-1}^{\dagger}\bi{a}_j^{} +
  2c_{2}(0)\,\bi{n}_{j}^{}\bi{n}_{j-1}^{} - 2c_{2}(0)\,\bi{n}_{j}^{}, \\
\fl \bi{H}_{1}^{\rm (b)} & = &
  -\frac{s_{2}(0)}{s_{0}(\psi_{-})}
  \left[c_{0}(\psi_{-})\,\bi{n}_{1}^{}-i\alpha_{-}\,\bi{a}^{\dagger}_{1}
        - i\beta_{-}\,\bi{a}_{1}^{}\right], \label{bt1}\\
\fl \bi{H}_{N}^{\rm (b)} & = &
  -\frac{s_{2}(0)}{s_{0}(\psi_+)}
  \left[c_{0}(\psi_{+})\,\bi{n}_{N}^{}-i\alpha_{+}\,\bi{a}^{\dagger}_{N}
        - i\beta_{+}\,\bi{a}_{N}^{}\right], \label{btN} \\
\fl \bi{H}_{m,m-1,m-2}^{\rm (h)} & = &
  \left\{\frac{s_{2}(0)-s_{2}(\nu_m)}{s_{2}(\nu_m)}
  \left[\bi{a}_{m-1}^{\dagger}\bi{a}_{m-2}^{}+\bi{a}_{m}^{\dagger}
        \bi{a}_{m-1}^{}\right]
\right.\nonumber \\
  \fl
  & &\left.+ \frac{s_{0}(\nu_m)}{s_{-2}(\nu_m)} \bi{a}^{\dagger}_{m-2}\bi{a}_{m}^{}\right\}+
       \mbox{\rm h.c.}, \label{im1}\\
\fl \bi{H}_{m,m-1,m-2}^{\rm (d)} & = &
  \frac{2c_{2}(0) s^{2}_{0}(\nu_m)}{\Delta(\nu_m)}
  \left[\bi{n}_{m}^{}\bi{n}_{m-2}^{}-\bi{n}_{m}^{}\bi{n}_{m-1}^{}-
        \bi{n}_{m-2}^{}\bi{n}_{m-1}^{}+
   \bi{n}_{m-1}^{}\right],\\
\fl \bi{H}_{m,m-1,m-2}^{\rm (c)} & = &
  \left\{\frac{s_{0}(\nu_m)s_{4}(0)}{\Delta(\nu_m)}
  \left[\bi{n}_{m}^{} \bi{a}_{m-2}^{\dagger}\bi{a}_{m-1}^{}
        - \bi{n}_{m-2}^{}\bi{a}_{m}^{\dagger}\bi{a}_{m-1}^{}\right]\right.\nonumber\\
  \fl & & \left. -\frac{2c_{2}(0) s^{2}_{0}(\nu_m)}{\Delta(\nu_m)}
    \bi{n}_{m-1}^{}\bi{a}_{m-2}^{\dagger}\bi{a}_{m}^{}\right\}\; +\; {\rm h.c.} ,
  \label{im2}
\end{eqnarray}
\endnumparts%
where
\begin{equation}
\Delta(u) \equiv s_{2}(u) s_{-2}(u)
\end{equation}
and h.c.\ denotes the hermitian conjugate with $(\psi_{\pm
  })^{*}=-\psi_{\pm }, \,\eta ^{*}=-\eta,\,\nu ^{*}=\nu$.  Here, $\bi
{H}^{\rm (b)}_{1}$ and $\bi {H}^{\rm (b)}_{N}$ are the general
boundary terms which are responsible for the sources and sinks with
particle injection and ejection at the boundaries; $\bi {H}^{\rm
  (h)}_{m,m-1,m-2}$ consists of nearest- and next-nearest-neighbour
hopping terms involving the sites $m$, $m-1$ and $m-2$; $\bi {H}^{\rm
  (d)}_{m,m-1,m-2}$ contains an onsite potential and density-density
interaction terms between neighbours and next-nearest neighbours; and
$\bi {H}^{\rm (c)}_{m,m-1,m-2}$ involves current-density interactions
(see \fref{fig-bmi}). The Hamiltonian in the presence of more than one
impurity can easily be constructed, if the two nearest impurities are
still well separated. In this case, the Hamiltonian reduces to a sum
over the isolated impurities like in the case of the Heisenberg
periodic chain \cite{shift,Schmitteckert,EPR}.

\section{Integrable impurities coupled to the boundaries}
\label{sec4}

\begin{figure}
\centerline{\psfig{file=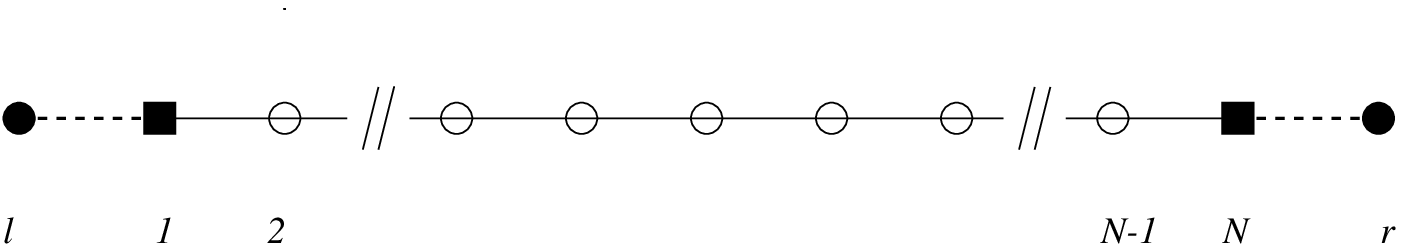,width=\figwidth}}
\caption{\label{fig-im+bmi}Impurities coupled to each of the
boundaries.}
\end{figure}

Kondo-like impurities of local impurity spins coupled to 1D strongly
correlated conduction electrons have attracted much interest
\cite{ref2,bi} especially in the context of the BA solution
\cite{Andrei}.  To every complex-valued $K$-matrix solution of the RE
\eref{3a} and \eref{3b}, one may construct a class of ``regular''
solutions \cite{reg}, i.e.,
\begin{equation}
\tilde{\bi K}_{-}(u)   =  \bi{L}(u)K_{-}(u)\bi{L}^{-1}(-u),\quad
\tilde{\bi K}_{+}(u)   =  K_{+}(u)
\end{equation}
to the same RE.  In order to study Kondo impurities for 1D electron
systems \cite{bi,reg}, it is better to construct ``non-regular'',
i.e.\ non-trivial operator-valued, solutions, to the RE.  Wang and
coworkers \cite{ref2w} constructed a class of integrable impurities
coupled to each boundary of the spin-$\frac{1}{2}$ Heisenberg XXZ
chain by a special choice of boundary $K_{\pm }$-matrices, i.e.,
$K_{\pm }=1$. But in their approach the parameters characterizing the
strength of the magnetic impurities --- related to our potential
impurities via the customary Jordan-Wigner transformation
\cite{Mattis} --- disappear in the Hamiltonian as well as in the BA
equations. Here we present a different approach to integrable
impurities: from ``regular'' solutions of the graded RE \eref{3a} and
\eref{3b}, we construct a class of integrable impurities
\cite{shift,Schmitteckert,EPR} coupled to each of the boundaries of a
fermion chain with general open BC.  We stress that these impurities
are not Kondo-like. If we embed two fermionic impurity vertices at the
boundaries,
\numparts%
\begin{eqnarray}
\fl
\bi{T}(u) & = & \bi{L}_{r}(u+\nu_{r})\bi{L}_{N}(u)\cdots \bi{L}_{m}(u)\cdots
\bi{L}_{1}(u)\bi{L}_{\ell}(u+\nu_{\ell}),\\
\fl
\bi{T}^{-1}(-u) & = & \bi{L}_{\ell}^{-1}(-u+\nu_{\ell})
\bi{L}_{1}^{-1}(-u)\cdots \bi{L}_{m}^{-1}(-u)
\cdots \bi{L}_{N}^{-1}(-u)\bi{L}_{r}^{-1}(-u+\nu_{r}),
\end{eqnarray}
\endnumparts%
one can show that
\begin{equation}
\bi{U}_{-}(u)=\bi{T}(u)K_{-}(u)\bi{T}^{-1}(-u)\label{u}
\end{equation}
also satisfies \eref{3a} and so does the solution
$\bi{L}_{\ell}(u+\nu_{\ell})K_{-}(u)\bi{L}^{-1}_{\ell}(-u+\nu_{\ell})$. It
follows that there exists a family of transfer matrices
\begin{equation}
\btau (u)={\rm Str}_{0}[K_{+}(u)\bi{U}_-(u)]\label{transf}
\end{equation}
and its members commute with each other for different spectral parameters.
Similarly to \eref{relat}, we can formulate the explicit expression of the
Hamiltonian for an open fermion chain with boundary impurities,
\begin{eqnarray}
\fl
\bi{H} & = & \sum_{j=2}^{N} \bi{H}_{j,j-1} +
        \frac{1}{{\rm Str}_{0}[K_{+}(0)]}
        \left\{ {\rm Str}_{0}[K_{+}(0)\bi{L}_{r}^{\prime}(\nu_{r})
                 \bi{L}_{r}^{-1}(\nu_{r})]
        \right.\nonumber \\
\fl
  &   & \hphantom{\sum_{j=2}^{N} H_{j,j-1}+
                  \frac{1}{{\rm Str}_{0}[K_{+}(0)]}\{}
         \left. + {\rm Str}_{0}[K_{+}(0)\bi{L}_{r}(\nu_{r})
         \bi{L}^{\prime}_{N}(0)\bi{L}_{N}^{-1}(0)\bi{L}_{r}^{-1}(\nu_{r})]
         \right\}\nonumber \\
\fl
  &   & + \frac{1}{2} \bi{L}_{1}(0) \bi{L}_{\ell}(\nu_{\ell})
          K_{-}^{\prime}(0)
          \bi{L}_{\ell}^{-1}(\nu_{\ell}) \bi{L}_{1}^{-1}(0)
        + \bi{L}_{1}(0) \bi{L}_{\ell}^{\prime}(\nu_{\ell})
          \bi{L}_{\ell}^{-1}(\nu_{\ell})
          \bi{L}_{1}^{-1}(0).
\label{for}
\end{eqnarray}
Substituting \eref{3e} and \eref{3f} into \eref{for},
the corresponding Hamiltonian is given as
\begin{eqnarray}
 \bi{H} & = &
\sum_{j=2}^{N} \bi{H}_{j,j-1} +
 \frac{s_{2}(0)}{\Delta(\nu_{r})s_{0}(\psi_{+})}
 \left[\bi{H}^{\rm (b)}_{N} + \bi{H}^{\rm (b)}_{r} +
       \bi{H}^{\rm (i)}_{N,r}\right] \nonumber\\
 & &
+ \frac{s_{2}(0)}{\Delta(\nu_{\ell})s_{0}(\psi_{-})}
\left[\bi{H}^{\rm (b)}_{1}+\bi{H}^{\rm (b)}_{\ell}+
      \bi{H}^{\rm (i)}_{1,\ell}\right] \label{4a}
\end{eqnarray}
where
\numparts%
\begin{eqnarray}
\fl \bi{H}^{\rm (b)}_{N} & = &
  s_{0}(\nu_r)\left[s_{0}(\psi_{+}-\nu_r)\,\bi{n}_{N}^{}
  +is_{2}(\nu_{r})\alpha_{+}\,\bi{a}^{\dagger}_{N}
  +is_{-2}(\nu_{r})\beta_{+}\,\bi{a}_{N}^{}\right]\nonumber\\
\fl & &
  - c_{2}(\nu_{r})s_{-2}(\nu_{r})s_{0}(\psi_{+})\,\bi{n}_{N}^{},\label{Ha}\\
\fl \bi{H}^{\rm (b)}_{r} & = &
  s_{2}(0) \left[ s_{2}(\psi_{+})\,\bi{n}_{r}^{}
  -is_{2}(\nu_{r})\alpha_{+}\,\bi{a}^{\dagger}_{r}
  +is_{-2}(\nu_{r})\beta_{+}\,\bi{a}_{r}^{}\right],\\
\fl
\bi{H}^{\rm (i)}_{N,r} & = &
  -s_{2}(0)c_{2}(0)\left[s_{0}(\psi_{+})\,\bi{n}_r^{}
  -2is_{0}(\nu_{r})(\alpha_{+}\,\bi{a}_{r}^{\dagger}-
  \beta_{+}\,\bi{a}_{r}^{})\right]
  \bi{n}_{N}\nonumber\\
\fl & &
  -c_{2}(0) \left[s_{2}(0) s_{0}(\psi_{+}) \,\bi{n}_{N}
  +2is^{2}_{0}(\nu_{r})(\alpha_{+}\,\bi{a}_{N}^{\dagger}
  +\beta_{+}\,\bi{a}_{N})\right] \bi{n}_{r}^{},\nonumber \\
\fl & &
  -s_{2}(0) \left[s_{0}(\nu_{r}+\psi_{+})\,\bi{a}_{N}^{\dagger}\bi{a}_r^{}
  -s_{0}(\nu_{r}-\psi_{+})\,\bi{a}_{r}^{\dagger}\bi{a}_{N}\right],\\
\fl
\bi{H}^{\rm (b)}_{1} & = &
\bi{H}^{\rm (b)}_{N}(N\rightarrow 1, r\rightarrow\ell,
\psi_{+}\rightarrow\psi_{-},\alpha_{+}\rightarrow\alpha_{-},
\beta_{+}\rightarrow\beta_{-}, \nu_{r}\rightarrow -\nu_{\ell}),\label{He}\\
\fl
\bi{H}^{\rm (b)}_{\ell} & = &
\bi{H}^{\rm (b)}_{r}(N\rightarrow 1, r\rightarrow\ell,
\psi_{+}\rightarrow\psi_{-},\alpha_{+}\rightarrow\alpha_{-},
\beta_{+}\rightarrow\beta_{-}, \nu_{r}\rightarrow -\nu_{\ell}),\label{Hf}\\
\fl
\bi{H}^{\rm (i)}_{1,\ell} & = &
\bi{H}^{\rm (i)}_{N,r}(N\rightarrow 1, r\rightarrow\ell,
\psi_{+}\rightarrow\psi_{-}, \alpha_{+}\rightarrow\alpha_{-},
\beta_{+}\rightarrow\beta_{-}, \nu_{r}\rightarrow -\nu_{\ell}),\label{Hg}
\end{eqnarray}
\endnumparts%
where $\bi {H}^{\rm {(b)}}$ can be interpreted as fermion sources
and sinks with particle injection and ejection at the boundaries and
at the impurity sites.  However, unlike the previous Hamiltonian
\eref{3g}, the boundary parameters and impurity parameters are both
involved.  $\bi{H}^{\rm (i)}$ describes the interaction between
impurities and boundaries (see \fref{fig-im+bmi}).

\begin{figure}
\centerline{\psfig{file=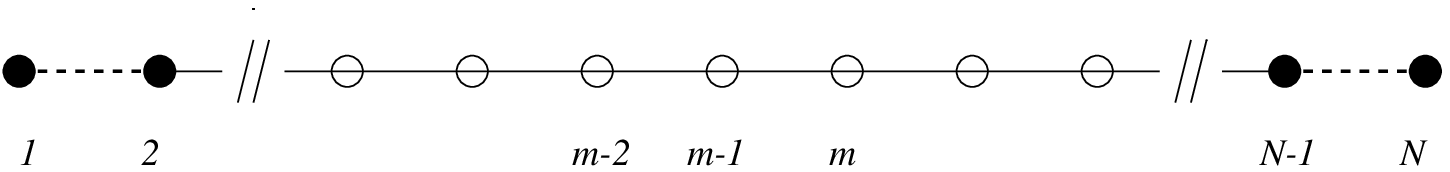,width=\figwidth}}
\caption{\label{fig-icbi}Integrable impurities situated at the
  boundaries.}
\end{figure}

\begin{figure}
\centerline{\psfig{file=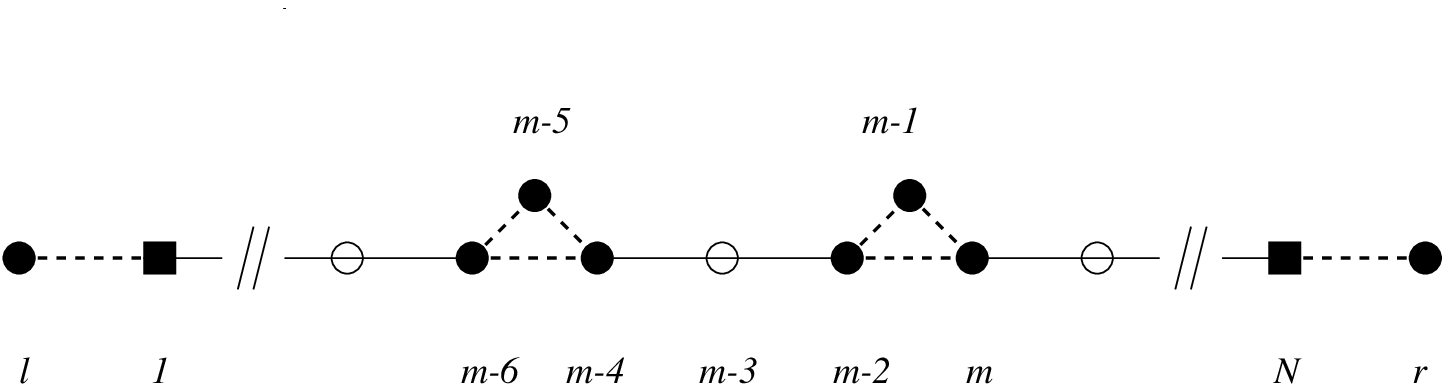,width=\figwidth}}
\caption{\label{fig-wsibmi} 
  Two well-separated bulk impurities at sites $m_1 = m$ and $m_2= m-4$
  together with the boundary impurities.}
\end{figure}

On the other hand, if we move the impurity in the bulk to each boundary
of the chain as shown in \fref{fig-icbi} with the monodromy matrix 
\numparts%
\begin{eqnarray}
\bi{T}(u) & = &
  \bi{L}_{N}(u+\nu_N)\cdots \bi{L}_m(u)\cdots \bi{L}_1(u+\nu_1), \\
\bi{T}^{-1}(-u) & = &
  \bi{L}^{-1}_1(-u+\nu_1)\cdots \bi{L}^{-1}_m(-u)\cdots
  \bi{L}^{-1}_N(-u+\nu_N),
\end{eqnarray}
\endnumparts%
one finds that the Hamiltonian is same as \eref{4a} apart from the
numbering
$$r\rightarrow N, N\rightarrow N-1, \ell\rightarrow 1, 1\rightarrow
2.$$ Although the eigenvalues of the open chain do not depend on the
position of the impurities in the bulk due to the absence of
back-scattering, the open boundary is a perfect back-scatterer with
vanishing transmission at each end of the open chain for
$\alpha_{\pm}=\beta_{\pm}=0$. 
Moreover, it is easy to obtain a model with the impurities coupled to
each boundary together with $f$ well separated impurities (see
\fref{fig-wsibmi}) at positions $m_i$ for $i=1, \ldots, f$, i.e.
\begin{eqnarray}
 \bi{H} & = &
\sum_{j=2}^{N} \bi{H}_{j,j-1} +
 \frac{s_{2}(0)}{\Delta(\nu_{r})s_{0}(\psi_{+})}
 \left[\bi{H}^{\rm (b)}_{N} + \bi{H}^{\rm (b)}_{r} +
       \bi{H}^{\rm (i)}_{N,r}\right] \nonumber\\
 & &
+ \frac{s_{2}(0)}{\Delta(\nu_{\ell})s_{0}(\psi_{-})}
\left[\bi{H}^{\rm (b)}_{1}+\bi{H}^{\rm (b)}_{\ell}+
      \bi{H}^{\rm (i)}_{1,\ell}\right] \nonumber\\
& &
+\sum_{i=1}^{f} (\bi{H}_{m_i,m_{i}-1,m_{i}-2}^{\rm (h)} + \bi{H}_{m_{i},m_{i}-1,m_{i}-2}^{\rm (d)}
+ \bi{H}_{m_{i},m_{i}-1,m_{i}-2}^{\rm (c)}).\label{4b}
\end{eqnarray}
The terms $\bi {H}^{\rm (b)},\,\bi {H}^{\rm (i)}$ are the same as in
\eref{Ha}-\eref{Hg}, and the terms $\bi {H^{\rm (c)}},\bi {H^{\rm
    (d)}}$ and $\bi {H^{\rm (h)}}$ are given in
\eref{im1}--\eref{im2}. To keep the Hamiltonians \eref{3g},
\eref{4a} and \eref{4b} hermitian, the parameters $\eta$ and
$\psi_{\pm}$ must be purely imaginary, $\nu $ real and $\alpha_{\pm
  }^{\dagger}=\beta_{\pm }$. All terms in the Hamiltonians are needed
to ensure the integrability of the models. In the next section we
shall proceed with the algebraic solutions for the small-polaron model
with different kinds of impurities in the most interesting special
case of perfectly back-scattering boundaries without sources and
sinks.

\section{Bethe ansatz solution for finite chains}
\label{sec5}

Following the method of \cite{Sklyanin,gfy}, we shall study the
algebraic BA solutions for the open fermion chain with different kinds
of impurities.  We first note that the general open BC spoil the
pseudo vacuum state.  Therefore it seems difficult to solve the models
with general open BC by means of the QISM. We thus restrict ourselves
to the simpler situation $\alpha_{\pm}=0$, $\beta_{\pm}=0$ in the
following. In this case, the Hamiltonians \eref{3g}, \eref{4a} and
\eref{4b} do not contain any Grassmannian source and sink terms and
are charge conserving. They still contain the potential impurities and
are perfect back-scatterers. Thus these Hamiltonians are ideal for the
proposed investigation of the interplay of forward-scattering bulk
impurities with backward-scattering boundaries.

Let us for simplicity first consider the Hamiltonian \eref{4a}.  In
the case $\alpha_{\pm}=0$, $\beta_{\pm}=0$, the Hamiltonian \eref{4a}
comprises
\numparts%
\begin{eqnarray}
\fl
\bi{H}^{\rm (b)}_{N} & = &
\left[s_{2}(0)c_{2}(0)s_{0}(\psi_+ )
-s_{0} ^2(\nu_{r})c_{0}(\psi_+)\right]\bi{n}_N,\label{5a-1}\\
\fl
\bi{H}^{\rm (b)}_{r} & = &
\left[s_{2}(0)c_{2}(0)s_{0}(\psi_+ )
+s_{2} ^2(0)c_{0}(\psi_+) \right]\bi{n}_{r},\\
\fl
\bi{H}^{\rm (i)}_{N,r} & = &
- s_{2}(0)
\left[
s_{0}(\nu_r+\psi_+)\,\bi{a}_{N}^{\dagger }\bi{a}_{r}
 - s_{0}(\nu_r-\psi_+)\,\bi{a}_{r}^{\dagger}\bi{a}_{N}
\right] - 
s_{4}(0)s_{0}(\psi_+)\bi{n}_{r}\bi{n}_{N}\label{5a-3}
\end{eqnarray}
\endnumparts%
and $\bi{H}_{1}^{(b)}$, $\bi{H}_{\ell}^{(b)}$, and
$\bi{H}_{1,\ell}^{(i)}$ follow as in \eref{He}--\eref{Hg}.  $\bi
{H^{{\rm (b)}}}$ describes the boundary impurities. $\bi
{H^{{\rm (i)}}}$ contains the interaction terms with exchange coupling
between the bulk and the impurities (see \fref{fig-im+bmi}). As
mentioned before, this Hamiltonian conserves the particle number due
to the absence of sources and sinks with particle injection and
ejection at the boundaries.

Now we proceed to establish the Bethe eigenvectors for the Hamiltonian
\eref{4a} with \eref{5a-1}-\eref{5a-3} by means of the algebraic BA
\cite{QISM}. Let
\begin{equation}
\bi {T}(u)  =
\left(\matrix{A&B\cr C&D\cr }\right),\quad
\bi {T}^{-1}(-u)  =
\left(\matrix{\bar{A}&\bar{B}\cr \bar{C}&\bar{D}}\right),
\end{equation}
be the monodromy matrices acting on the pseudo vacuum state defined by
$\bi{a}_j|0\rangle =0,\,j=1,\cdots ,N$. Then we have
\numparts%
\begin{eqnarray}
A|0\rangle & = & s_{0}^{N}(u)s_{0}(u+\nu_{\ell})s_{0}(u+\nu_{r})|
0\rangle,\label{va}\\
D|0\rangle & = & s_{2}^{N}(u)s_{2}(u+\nu_{\ell })s_{2}(u+\nu
_{r})|0\rangle,\\
B| 0 \rangle & = & 0,\\
C| 0\rangle & \neq & 0,\\
\bar{A}| 0 \rangle & = &\frac{(-1)^Ns_{0}^{N}(u)s_{0}(u-\nu_{\ell})s_{0}(u-\nu
_{r})} {\delta\left[\bi {T}(-u)\right]}| 0\rangle , \\
\bar{D}| 0 \rangle & = &\frac{(-1)^Ns_{2}^{N}(u)s_{2}(u-\nu
_{\ell})s_{2}(u-\nu_{r})}{\delta\left[\bi {T}(-u)\right]}| 0\rangle
,\\
\bar{B}| 0\rangle  & = & 0,\\
\bar{C}| 0\rangle &\neq & 0,\label{vb}
\end{eqnarray}
\endnumparts%
where the quantum determinant \cite{KS} is $\delta\left[\bi
  {T}(-u)\right] = \Delta ^{N}(u )\Delta (u-\nu_{\ell} )\Delta
(u-\nu_{r})$. Let us define
\begin{equation}
  \bi {U}_-(u)=\left(\matrix{\tilde A&\tilde B\cr \tilde C&\tilde D\cr
      }\right).\label{5b}
\end{equation}
From \eref{u}, we then have
\numparts%
\begin{eqnarray}
\tilde A & = &\frac{1}{s_{0}( \psi_-)}[-s_{0}(u-\psi
_-)A\bar{A}+s_{0}(u+\psi_-)B\bar{C}],\\
\tilde D & = &\frac{1}{s_{0}
(\psi_-)}[-s_{0}(u-\psi_-)C\bar{B}+s_{0}(u+\psi_-)D\bar{D}].
\end{eqnarray}
\endnumparts%
Noting the following form of the graded YBA
\begin{equation}
 \stackrel{2}{\bi{T}}\mbox{$\!$}^{-1}(-u)
 R_{12}(2u)
 \stackrel{1}{\bi{T}}\!(u) =
 \stackrel{1}{\bi{T}}\!(u)
 R_{12}(2u)
 \stackrel{2}{\bi{T}}\mbox{$\!$}^{-1}(-u),
\end{equation}
it is possible to derive the commutation relation
\begin{equation}
B\bar{C}=\frac{s_{2}(0)}{s_{2}(2u)}(\bar{D}D-A\bar{A}).
\end{equation}
With the help of the graded RE \eref{3a} we obtain --- after a lengthy
calculation --- the commutation relations
\numparts%
\begin{eqnarray}
\fl
\hat{A}(u)\tilde C(v)
& = &
\frac{s_{2}(u-v)s_{4}(u+v)}{s_{0}(u-v)s_{2}(u+v)}
\tilde C(v)\hat{A}(u)
-\frac{s_{2}(0)s_{4}(2u)}{s_{2}(2u)s_{0}(u-v)}
\tilde C(u)\hat{A}(v) \nonumber \\
\fl
& &
+\frac{s_{2}(0)s_{0}(2v)s_{4}(2u)}{s_{2}(2v)s_{2}(2u)
s_{2}(u+v)}\tilde C(u)\tilde D(v),\label{com1}\\
\fl
\tilde D(u)\tilde C(v)
& = &
\frac{s_{0}(u+v)s_{-2}(u-v)}{s_{0}(u-v)s_{2}(u+v)}
\tilde C(v)\tilde D(u)
+\frac{s_{2}(0)s_{0}(2v)}{s_{0}(u-v)s_{2}(2v)}
\tilde C(u)\tilde D(v)\nonumber \\
\fl
& & -\frac{s_{2}(0)}{s_{2}(u+v)}\tilde C(u)
\hat{A}(v),\label{com2}
\end{eqnarray}
\endnumparts%
where we introduced the transformation
\begin{equation}
\hat{A}(u)=\tilde A(u)-\frac{s_{2}(0)}{s_{2}(2u)}\tilde
D(u).
\end{equation}
{}From \eref{va}-\eref{vb} and \eref{5b}, we can choose an $M$-particle
excitation as
\begin{equation}
| \Phi (v_1\cdots v_M)\rangle =\tilde C(v_1)\cdots \tilde
C(v_M)| 0\rangle .
\end{equation}
Using the commutation relations \eref{com1} and \eref{com2}, one
obtains the eigenvalue $\Lambda$ of the transfer matrix \eref{transf}
\begin{equation}
  \tau (u)| \Phi (v_1\cdots v_M)\rangle =\Lambda (u;v_1\cdots
  v_M)| \Phi (v_1\cdots v_M)\rangle ,
\end{equation}
where
\begin{eqnarray}
\fl
\Lambda (u;v_1\cdots v_M)  =
- \frac{(-1)^N}{s_{2}(2u)\delta
 \left[\bi {T}(-u)\right]s_{0}(\psi_-)}  \nonumber \\
{\times}  \left\{\rule{0em}{4ex}
s_{2}(u-\psi_+)s_{2}(u-\psi_-)
s_{0}(u-\nu_{\ell})s_{0}(u+\nu_{\ell})
s_{0}(2u)s_{0}^{2N}(u)
\right. \nonumber \\
 \left.
{\times} s_{0}(u-\nu_{r})s_{0}(u+\nu_{r})\prod
_{j=1}^M\frac{s_{4}(u+v_j)s_{2}(u-v_j)}{s_{0}
  (u-v_j)s_{2}(u+v_j)}
\right. \nonumber \\
\left.
+  s_{0}(u+\psi_+)s_{0}(u+\psi_-)s_{2}(u+\nu
_{\ell})
s_{2}(u-\nu_{\ell})s_{4}(2u)s_{2}^{2N}(u)
\right. \nonumber \\
 \left. {\times}
s_{2}(u+\nu_{r})
s_{2}(u-\nu_{r})\prod_{j=1}^{M}\frac{s_{0}(u+v_j)s_{-2}(u-v_j
)}{s_{0}(u-v_j)s_{2}(u+v_j)}
\right\} , \label{4aa}
\end{eqnarray}
provided that
\begin{eqnarray}
\fl
\frac{s_{1}(v_j-\psi_-)s_{1}(v_j-\psi_+)s_{-1}
^{2N}(v_j)} {s_{-1}(v_j+\psi_-)s_{-1}(v_j+\psi
_+)s_{1}^{2N}(v_j)} = \nonumber\\
 \mbox{ } \prod_{m=\ell,r}\frac{s_{1}(v_j+\nu_m)s_{1}
  (v_j-\nu_m)}{s_{-1}(v_j+\nu_m)s_{-1}(v_j-\nu
_m)}
\prod_{\mbox{\scriptsize $\begin{array}{c}k=1\\ k\neq
  j\end{array}$}}^{M}\frac{s_{-2}(v_j+v_k
  )s_{-2}(v_j-v_k)}{s_{2}(v_j+v_k)s_{2}(v_j-v_k
  )}\label{5c}
\label{eq:fs-ba-rl}
\end{eqnarray}
for all $j=1,\ldots, M$. In the above BA equations, we have shifted the
parameter $v_j\rightarrow v_j-\eta $.  From the relation \eref{relat},
the eigenvalue $E$ of the Hamiltonian \eref{4a} for
$\alpha_{\pm}=\beta_{\pm}=0$  follows as
\begin{eqnarray}
\fl
E  =  -s_{2}(0)\left[\rule{0em}{4ex}\cot \psi_-+\cot \psi_++2(N+1)\cot 2\eta
-2\cot (\nu_{r}-2\eta ) -2\cot (\nu_{\ell}-2\eta) \right. \nonumber \\
\left. 
-\sum_{j=1}^{M}\frac{2s_{2}(0)}{s_{-1}(v_j)s_{1}(v_j
)}\right].\label{5ce}
\label{eq:fs-energy-rl}
\end{eqnarray}
On the right hand side of \eref{5ce} we have dropped a multiplicative
term $1/2 \cos(2\eta)$ as is customary \cite{Sklyanin}. We emphasize
that the spatial position of the impurities in the chain, although
clearly important in the construction of the Hamiltonian, is
irrelevant for the BA equations and the ground state energy
\cite{shift,Schmitteckert,EPR}. This is the mathematical formulation
of the physical absence of backscattering for these impurities
\cite{EPR}.

The BA equations and the eigenvalues for
the Hamiltonian \eref{3g} can be obtained similarly as, 
\begin{eqnarray}
\fl
\frac{s_{1}(v_j-\psi_-)s_{1}(v_j-\psi_+)s_{-1}
^{2(N-1)}(v_j)} {s_{-1}(v_j+\psi_-)s_{-1}(v_j+\psi_+)s_{1}^{2(N-1)}(v_j)}
= \nonumber\\
 \frac{s_{1}(v_j+\nu_m)s_{1}
  (v_j-\nu_m)}{s_{-1}(v_j+\nu_m)s_{-1}(v_j-\nu_m)}
\prod_{\mbox{\scriptsize $\begin{array}{c}k=1\\ k\neq
  j\end{array}$}}^{M}\frac{s_{-2}(v_j+v_k )s_{-2}(v_j-v_k)}
  {s_{2}(v_j+v_k)s_{2}(v_j-v_k)}
\label{eq:fs-ba-f}
\end{eqnarray}
for all $j=1,\ldots, M$, and
\begin{equation}
\fl
 E  =  -s_{2}(0)\left[\cot \psi_- +\cot \psi_+ + 2N \cot 2\eta
-2\cot (\nu_m-2\eta ) 
 -\sum_{j=1}^{M}\frac{2s_{2}(0)}{s_{-1}(v_j)s_{1}(v_j)}\right].
\label{eq:fs-energy-f}
\end{equation}
For the Hamiltonian \eref{4b} we find
\begin{eqnarray}
\fl
\frac{s_{1}(v_j-\psi_-)s_{1}(v_j-\psi_+)s_{-1}
^{2(N-f)}(v_j)} {s_{-1}(v_j+\psi_-)s_{-1}(v_j+\psi
_+)s_{1}^{2(N-f)}(v_j)} = \nonumber\\
 \mbox{ } \prod^{f,\ell,r}_{m=1}\frac{s_{1}(v_j+\nu_m)s_{1}
  (v_j-\nu_m)}{s_{-1}(v_j+\nu_m)s_{-1}(v_j-\nu
_m)}
\prod_{\mbox{\scriptsize $\begin{array}{c}k=1\\ k\neq
  j\end{array}$}}^{M}\frac{s_{-2}(v_j+v_k
  )s_{-2}(v_j-v_k)}{s_{2}(v_j+v_k)s_{2}(v_j-v_k
  )},
\label{eq:fs-ba-frl}
\end{eqnarray}
where $\prod_{m=1}^{f,\ell,r}$ denotes the product over the $f$ isolated
impurities in the bulk as well as the boundary impurities.  The energy
spectrum is given as
\begin{eqnarray}
\fl
E = -s_{2}(0)\left[\rule{0em}{4ex}\cot \psi_-+\cot \psi_++2(N-f+1)\cot 2\eta
-2\sum_{m=1}^{f,\ell,r}\cot (\nu_m-2\eta )
\right. \nonumber\\
 \left.  -\sum
_{j=1}^{M}\frac{2s_{2}(0)}{s_{-1}(v_j)s_{1}(v_j)}\right].
\label{eq:fs-energy-frl}
\end{eqnarray}
When comparing \eref{eq:fs-ba-frl}, \eref{eq:fs-energy-frl} to
\eref{eq:fs-ba-f}, \eref{eq:fs-energy-f} and \eref{eq:fs-ba-rl},
\eref{eq:fs-energy-rl}, we see that using \eref{eq:fs-ba-frl},
\eref{eq:fs-energy-frl} corresponding to the Hamiltonian \eref{4b}, we
can reproduce the results for the other two Hamiltonians \eref{3g} and
\eref{4a}. Namely, with $f=1$ and $\nu_r=\nu_l=0$ we get the result
for \eref{3g} with \eref{eq:fs-ba-f}, \eref{eq:fs-energy-f} and with
$f=0$ and $\nu_r, \nu_l\neq 0$ we find for \eref{4a} the BA equations
\eref{eq:fs-ba-rl} with energy \eref{eq:fs-energy-rl}. Thus the
Hamiltonian \eref{4b} contains the other two Hamiltonians as special
cases, although the construction by QISM proceeds independently. We
note that care has to be paid to the varying number of sites
$N$ when doing this procedure.

\section{Ground-state properties in the thermodynamic limit}
\label{sec6}

We note that the bulk terms of the Hamiltonians \eref{3g}, \eref{4a}
and \eref{4b} are equivalent to the counter part of the 1D Heisenberg XXZ
model with periodic BC via a Jordan-Wigner transformation. The
finite-size corrections and thermodynamics for the XXZ model with or
without boundary magnetic fields have been studied in many papers
\cite{ref12,ref13}. As mentioned before, the Jordan-Wigner
transformation does not preserve the boundary terms nor the
impurity terms due to its nonlocality. The BA equations we obtained
provide a more meaningful description of the ground-state
properties due to the presence of the boundary potential terms and the
impurity parameters. In what follows, we shall study the ground-state
properties for the resulting models following the scheme in
\cite{Schmitteckert,ref2w,ref12,ref13}.

For convenience, let us first redefine the variable $v_j\rightarrow
iv_j$. Then, taking the logarithm, we rewrite the BA equations
\eref{5c} for the Hamiltonian \eref{4a} as follows:
\begin{eqnarray}
\fl
2\pi I_j
 =
2N\theta (v_j,\eta)+\theta (v_j,\psi_+-\eta )+\theta(v_j,\psi_--\eta )
+\theta (v_j+\nu_{r},\eta)+\theta(v_j-\nu_{r},\eta)\nonumber \\
+\theta (v_j+\nu_{\ell},\eta)+\theta (v_j-\nu_{\ell},\eta)
-\sum_{\mbox{\scriptsize $\begin{array}{c}k=1\\ k\neq
  j\end{array}$}}^M \theta (v_j-v_k,2\eta)+\theta
(v_j+v_k,2\eta),\label{6a}
\end{eqnarray}
for all $j = 1,\cdots, M$, where the two-body phase shift
\cite{EK,YBE,QISM} is
\begin{equation}
  \theta (v_j,\eta)=i\ln \frac{\sinh
    (v_j+i\eta)}{\sinh(v_j-i\eta)}=2{\mathrm{arccot}} 
\left(\tanh v_j\cot\eta\right).
\end{equation}
We now define $v_{-k}$ as $-v_k$ and define $v_0=0$. Then the
density of the roots $\{v_j\}$ can be defined as
\begin{equation}
\rho_N (v)=\frac{dZ_N(v)}{dv}, \quad Z_N=\frac{I_j}{N}
\end{equation}
and the finite-size BA equation \eref{6a} becomes
\begin{equation}
\fl
Z_N(v)=\frac{1}{\pi }\left\{\theta (v, \eta )+\frac{1}{2N}\left[\theta
^{{\rm (i)}}(v)+\theta ^{{\rm (b)}}(v)\right]-\frac{1}{2N}\sum
_{k=-M}^M\theta (v-v_k,2\eta )\right\},\label{6b}
\end{equation}
where
\numparts%
\begin{eqnarray}
\theta ^{{\rm (i)}}(v) & = & \theta (v+\nu_{r},\eta )+\theta (v-\nu
_{r},\eta )+\theta (v+\nu_{\ell},\eta )+\theta (v-\nu_{\ell},\eta ),
\label{eq:thetai}\\
\theta ^{{\rm (b)}}(v) & = & \theta (v,2\eta )+\theta (2v,2\eta
)+\theta (v,\psi_+-\eta )+\theta (v,\psi_--\eta ).
\label{eq:thetab}
\end{eqnarray}
\endnumparts%
We note that the first two terms in \eref{eq:thetab} arise due to the
non-periodicity of the chain, whereas the last two terms are due to
the boundary potentials at sites $1$ and $N$.  Taking the thermodynamic
limit and differentiating \eref{6b} with respect to the spectral
parameter $v$, we have
\begin{equation}
\fl
\rho_{\infty}(v) = \frac{1}{\pi }\left\{ \theta^{\prime}(v, \eta
)+\frac{1}{2N}\left[{\theta ^{{\rm (i)}}}^{\prime}(v)+{\theta ^{{\rm (b)}}}^{\prime}(v)\right]\right\}
-\frac{1}{2\pi }\int_{-\Lambda }^{\Lambda }du\rho_{\infty}(u)\theta
^{\prime}(v-u,2\eta),\label{6c}
\end{equation}
where the integration boundary $\Lambda $ is determined by
\begin{equation}
\int_{-\Lambda }^{\Lambda }\rho_{\infty}(v)dv
=\frac{2M+1}{N}+O(N^{-2}).
\end{equation}
The prime denotes the derivative with respect to $ v $. Due to the
linearity of \eref{6c}, one may formally solve the following three
linear integral equations: 
\numparts%
\begin{eqnarray}
\rho^{(0)}_{\infty}(v) & = & \frac{1}{\pi }\theta ^{\prime}(v, \eta )
-\frac{1}{2\pi }\int_{-\Lambda }^{\Lambda }du\rho^{(0)}_{\infty}(u)\theta
^{\prime}(v-u,2\eta),\label{6d}\\
\rho ^{{\rm (i)}}_{\infty}(v) & = &
\frac{1}{\pi }{\theta ^{{\rm (i)}}}^{\prime}(v) -\frac{1}{2\pi }\int_{-\Lambda
}^{\Lambda }du\rho ^{{\rm (i)}}_{\infty}(u)\theta
^{\prime}(v-u,2\eta),\label{6e}\\
\rho ^{{\rm (b)}}_{\infty } (v) & = &
\frac{1}{\pi }{\theta ^{{\rm (b)}}}^{\prime}(v) -\frac{1}{2\pi }\int_{-\Lambda
}^{\Lambda }du\rho ^{{\rm (b)}}_{\infty } (u)\theta
^{\prime}(v-u,2\eta),\label{6f}
\end{eqnarray}
\endnumparts%
In this way, the solution of \eref{6c} can be expressed as
\begin{equation}
\rho_{\infty}(v)=\rho^{(0)}_{\infty}(v)+\frac{1}{2N}[\rho ^{{\rm
(i)}}_{\infty}(v)+\rho ^{{\rm (b)}}_{\infty } (v)],\label{6g}
\end{equation}
where $\rho^{(0)}_{\infty}(v),$ $\frac{1}{2N}\rho ^{{\rm
    (i)}}_{\infty}(v)$ and $\frac{1}{2N}\rho ^{{\rm
    (b)}}_{\infty}(v)$ are the contributions of the bulk, the
impurities and the boundary effect to the root density, respectively.
The ground-state energy \eref{5ce} is minimized at the cutoff $\Lambda
$ in the thermodynamic limit as discussed in \cite{ref12,ref13}.
Following the argument in \cite{ref12,ref13}, we find the cutoff
$\Lambda =\infty$ such that the particle density is $M/N=1/2$.

By using the Fourier transforms, the formal solutions to the equations
\eref{6d}--\eref{6f} read
\begin{equation}
\tilde{\rho}_{\infty}(\omega, \eta )=
\frac{2\tilde{\theta}(\omega, \eta)}{2\pi+\tilde{\theta}(\omega, 2\eta )},
\end{equation}
where
\begin{equation}
\tilde {\theta }(\omega ,\eta )=\int_{-\infty}^{\infty}\theta
^{\prime}(v,\eta )e^{i\omega v}dv.
\end{equation}
{}From the residue theorem, we obtain
\numparts%
\begin{eqnarray}
\rho^{(0)}_{\infty}(v) & = & \frac{2}{\eta \cosh \frac{\pi }{2\eta}v},\\
\rho ^{{\rm (i)}}_{\infty}(v) & = & \sum_{m=r,\ell}\frac{4\cosh \frac{\pi
}{2\eta }v\cosh \frac{\pi }{2\eta}\nu_m}{\eta \cosh \frac{\pi}{2\eta
}(v+\nu_m) \cosh \frac{\pi}{2\eta }(v-\nu_m)},\\
\rho ^{{\rm (b)}}_{\infty } (v) & = & \frac{1}{2\pi
}\int_{-\infty}^{\infty}
\left[
\tilde{\rho }^{{\rm(be)}}_{\infty}(\omega )
+\tilde{\rho }^{{\rm (bp)}}_{\infty }(\omega) 
\right] e^{-i\omega v} d\omega,\label{eq:rhob}
\end{eqnarray}
\endnumparts%
where
\numparts%
\begin{eqnarray}
\tilde{\rho}^{{\rm (be)}}_{\infty}(\omega ) & = &
\frac{2\sinh(\frac{\pi }{2}-2\eta)\omega +4\cos \frac{\pi 
}{4}\omega \sinh(\frac{\pi }{4}-\eta)\omega}{\sinh \frac{\pi
}{2}\omega +\sinh(\frac{\pi}{2}-2\eta )\omega },\\
\tilde{\rho }^{{\rm (bp)}}_{\infty}(\omega)& = &
\frac{2\sinh(\frac{\pi }{2}+\eta-\psi_+)\omega+
 2\sinh (\frac{\pi }{2}+\eta -\psi_-)\omega }
 {\sinh \frac{\pi }{2}\omega+\sinh(\frac{\pi}{2}-2\eta )\omega }.\label{bef}
\end{eqnarray}
\endnumparts%
Here $\tilde{\rho }^{{\rm (be)}}_{\infty}(\omega )$ and
$\tilde{\rho }^{{\rm (bp)}}_{\infty}(\omega)$ are the contributions to
the root density from the boundary effect and the boundary potential terms,
respectively, due to \eref{eq:thetab}.  Then, from \eref{5ce}, we
also obtain the ground-state energy in the thermodynamic
limit as
\begin{equation}
  E_{\rm g}=
  N\int_{-\infty}^{\infty}dv
  \frac{4\sin^22\eta}{\cosh 2v-\cosh 2\eta} \rho_{\infty}(v)
  + E_0,\label{6gr}
\end{equation}
with
\begin{equation}
E_0 = -p_+-p_--2(N+1)\cos 2\eta+2\sin 2\eta \sum_{m=r,\ell}\cot (\nu_m-2\eta
).
\end{equation}
The boundary energy \cite{ref13} is given by
\begin{eqnarray}
E_{\rm b} & = & \int_{-\infty}^{\infty}dv \frac{2\sin
^22\eta}{\cosh 2v-\cosh 2\eta}[\rho ^{{\rm (i)}}_{\infty}(v)+\rho ^{{\rm (b)}}_{\infty } (v)] \\
& &- p_+ - p_- - 2\cos 2\eta+2\sin 2\eta \sum_{m=r,\ell}\cot
(\nu_m-2\eta ). \nonumber
\end{eqnarray}
We thus note that the boundary potential terms do not only enter the
expression for the ground-state energy explicitly as $-p_+ -p_-$, but
also implicitly via $\tilde{\rho} ^{{\rm (b)}}_{\infty }$ of \eref{eq:rhob}.

We remark that in \cite{ref2w} boundary magnetic field terms,
corresponding to \eref{bef}, do not contribute to the root density due
to the lack of free boundary parameters in the boundary
$K_{\pm}$-matrices.  The presence of boundary potentials (magnetic
fields) and the impurity parameters changes the asymptotic behaviour
of the BA equations \eref{6a} resulting in string solutions different
from those discussed in the papers \cite{ref13}. Indeed, either the
boundary parameters $p_{\pm }$ (or $\psi_\pm $) or the impurity
strength parameters $\nu_{m}$, $\nu_{\ell}$ and $\nu_{r}$ affect the
boundary string solutions to the BA equations. It is obvious that the
ground-state energy of the bulk is same as in the periodic case
\cite{ref12}. In general, the boundary states are associated with
complex roots of the BA equations.

Analogously, we obtain the ground-state energy \eref{6gr} for the
Hamiltonians \eref{3g} and \eref{4b}. The differences to the
ground states for these Hamiltonians are only the contributions from
the impurities  expressed in $\rho ^{{\rm (i)}}_{\infty}(v)$. Thus
we get for the Hamiltonian \eref{3g}
\begin{eqnarray}
\rho ^{{\rm (i)}}_{\infty}(v) & =& \frac{4\cosh \frac{\pi }{2\eta }v\cosh
\frac{\pi
}{2\eta}\nu _m}{\eta \cosh \frac{\pi}{2\eta }(v+\nu _m) \cosh
\frac{\pi}{2\eta }(v-\nu _m)}- \frac{4}{\eta \cosh \frac{\pi }{2\eta
}v},\label{r1}
\\
E_0 & = & -p_+-p_--2N\cos 2\eta+2\sin 2\eta \,\cot (\nu _m-2\eta
).
\end{eqnarray}
For the most general Hamiltonian \eref{4b}, we have
\begin{eqnarray}
\rho ^{{\rm (i)}}_{\infty}(v) & = & \sum
_{m}^{f,r,\ell}\left[\frac{4\cosh \frac{\pi
}{2\eta }v\cosh \frac{\pi \nu
}{2\eta}\nu _m}{\eta \cosh \frac{\pi}{2\eta }(v+\nu _m) \cosh
\frac{\pi}{2\eta }(v-\nu _m)}\right]- f\frac{4}{\eta \cosh \frac{\pi
}{2\eta}v}. \label{r3}
\\
E_0 & = & -p_+-p_--2(N+1-f)\cos 2\eta \nonumber\\
& & \mbox{ }
+2\sin 2\eta \,\sum _{m=1}^{f,r,\ell}\cot (\nu _m-2\eta
).
\end{eqnarray}
Further thermodynamic properties such as compressibilities and
susceptibilities can be calculated as demonstrated previously in
\cite{shift,EPR,BCFT}. Results will be presented elsewhere.

\section{Conclusions and discussion}
\label{sec7}

In the present work, we have considered the interplay of integrable
impurities and general open boundary conditions for the example of the
small-polaron model. The impurities have been constructed via
inhomogeneous shifts in the spectral parameters of the Lax operators
such that the YBE are satisfied. The boundary terms are taken to obey
the RE. In both cases, we dealt with the graded version of the
equations due to the fermionic nature of the particles and the
boundary terms. Thus by construction, the model remains integrable.
We have shown that this is true even when placing the impurities
directly at the boundaries.

The most general boundary terms considered in \eref{2a} include
fermionic particle sources and sink terms as well as more standard
density terms.  However, these linear terms in creation and
annihilation operators contain coefficients that are odd Grassmann
variables. Thus a straightforward physical interpretation appears
problematic.  Representing these coefficients $\alpha_\pm$,
$\beta_\pm$ as additional fermionic operators $\bi{a}_{\pm}$,
$\bi{a}^{\dagger}_{\pm}$, we arrive at a chain with two additional
sites but without sources and sinks.

The boundary terms coupling to the particle density can be viewed as
potential impurities --- much like in the Anderson model of
localization \cite{KraM93} --- situated at the boundaries. For the
special case with only these potential impurities and the integrable
impurities present, we solve the BA equations and compute the
ground-state energy in the thermodynamic limit for half-filling. We
find that the solution is possible regardless whether the integrable
impurities are located within the bulk or at the boundaries.

The two types of impurities enter the expressions for the ground-state
energy additively. Thus the simultaneous presence of both purely
forward scattering integrable impurities and purely reflecting
boundary potential terms does not seem to change the physics in a
substantial way. We therefore do not expect to see the onset of
localization as might be expected from the form of the boundary
impurities.

An analogue of the integrable impurities can be found in the case of
light waves. Consider a long strip of glass, interspersed with pieces
of bifringent material of the same index of refraction as the glass.
Then as circular polarized light enters the strip and reaches the
first bifringent slab, its plane of polarization will be rotated by an
angle $\theta \propto \alpha l_1$ where $l_1$ represents the length of
the first slab and $\alpha$ is the material specific rotation angle
per unit length \cite{Hec87}. There is no reflection at the contact
due to the identical indices of refraction. After the next slab, we
have $\theta=\alpha l_1 + \alpha l_2$ and so on. Thus the net effect
of the bifringent slabs is a rotation of the plane of polarization,
i.e., a change in the overall phase of the wave function of light just
as for the electronic wave function in Ref.\ 
\cite{shift,Schmitteckert,EPR}. In this picture the integrable
boundaries satisfying the RE then simply correspond to perfect mirrors
at both ends of the strip.

\ack
The authors thank H.-P.\ Eckle for helpful discussions.  GXW thanks
H.-Q. Zhou and M.\ Shiroishi for their communications and also thanks
the Institut f\"{u}r Physik, Technische Universit\"{a}t Chemnitz for
kind hospitality. UG, RAR and MS gratefully acknowledge the
hospitality of the Max-Planck-Institut f\"{u}r Physik komplexer
Systeme (Dresden) during COOP99 where much of this manuscript was
prepared. The work has been supported by the DFG via SFB393 and the
Schwerpunktprogramm ``Quasikristalle''.



\Bibliography{99}

\bibitem{Mattis}
 Mattis D C 1993
 {\it The many-body problem: An Encyclopedia of Exactly Solved Models in One Dimension}
 (World Scientific, Singapore)

\bibitem{EK}
 Korepin V E and E{\ss}ler F H L 1994 
 {\it Exactly Solvable Models of Strongly Correlated Electrons}
 (World Scientific, Singapore)

\bibitem{Andrei}
 Andrei N, Furuya K and Lowenstein J H 1983 
 Solution of the Kondo problem 
 \RMP {\bf 55} 331--402
\item[]
 Tsvelik A M and Wiegmann P B 1983
 Exact results in the theory of magnetic alloys
{\it Adv.\ Phys.}\ {\bf 32} 453--713

\bibitem{Affleck}
 Affleck I 1990
  A current algebra approach to the Kondo effect
 {\it Nucl. Phys.} B {\bf 336} 517--32
\item[]
 Fr\"{o}jdh P and Johannesson H 1995
 Kondo effect in a Luttinger liquid: Exact results
 from conformal field theory
 \PRL {\bf 75} 300--3

\bibitem{Kondo} Kondo J 1964 
Resistance Minimum in Dilute Magnetic Alloys
{\it Progr. Theor. Phys.} {\bf 32} 37--49

\bibitem{spindefect}
 Schlottmann P 1991
 Impurity-induced critical behaviour in antiferromagnetic Heisenberg chains
 \JPCM {\bf 3} 6617--34
\item[]
Aladim S R and Martins M J 1993 Critical behaviour of integrable mixed-spin chains
\JPA {bf 26} L529-L534
\item[]
 de Vega H J and Woynarovich F 1992
 New integrable quantum chains combining different kinds of spins
 \JPA {\bf 25} 4499--516
\item[]
 de Vega H J, Mezincescu L and Nepomechie R I 1994
 Thermodynamics of integrable chains with alternating spins
 \PR B {\bf 49} 13223--6

\bibitem{AJ}
 Andrei N and Johannesson H 1984
Heisenberg chain with impurities \PL A {\bf 100} 108--12

\bibitem{shift}
 Bares P A 1994
 Exact results for a one-dimensional $t$-$J$ model with impurity
 {\it Preprint} cond-mat/9412011
\item[]
 Schmitteckert P, Schwab P and Eckern U 1995
 Quantum coherence in an exactly solvable one-dimensional model
 with defects
 {\it Europhys. Lett.} {\bf 30} 543--8

\bibitem{Schmitteckert}
 Schmitteckert P 1996
 {\it Interplay between Interaction and Disorder in One-dimensional
  Fermi Systems}
 Dissertation, University of Augsburg

\bibitem{EPR}
 Eckle H-P, Punnoose A and R\"{o}mer R A 1997
 Absence of backscattering at integrable impurities in one-dimensional
 quantum many-body systems
 {\it Europhys. Lett.} {\bf 39} 293--8

\bibitem{Sklyanin}
 Sklyanin E K 1988
 Boundary conditions for integrable quantum systems
 \JPA {\bf 21} 2375--89
\bibitem{bound}
 Cherednik I V 1984
 Factorizing particles on a half line and root systems
 {\it Theor. Math. Phys.} {\bf 61} 977--83
 [{\it Teor. Mat. Fiz.} {\bf 61} 35--44]
\bibitem{YBE}
Yang C N 1967 Some exact results for the many-body problem in
one dimension with repulsive delta-function interaction  \PRL {\bf 19} 1312--4
\item[]
 Baxter R J 1982
 {\it Exactly Solved Models in Statistical Mechanics}
 (London: Academic Press)
\bibitem{QISM}
 Korepin V E, Bogoliubov N M and Izergin A G 1993
 {\it Quantum Inverse Scattering Method and Correlation Functions}
 (New York: Cambridge University Press)

\bibitem{BCFT}
 Cardy J L 1989
 Boundary conditions, fusion rules and the Verlinde formula
 {\it Nucl. Phys.} B {\bf 324} 581--96
\bibitem{ref1}
 Frahm H and Zvyagin A A 1997
 The open spin chain with impurity: an exact solution
\JPCM {\bf 9} 9939--46
\item[]
 Bed\"{u}rftig G and Frahm H 1997
 Spectrum of boundary states in the open Hubbard chain
\JPA {\bf 30} 4139--49
\item[]
 Bed\"{u}rftig G, Brendel B, Frahm H and Noack R M 1998
 Friedel oscillations in the open Hubbard chain
 \PR  B {\bf 58} 10225--35
\bibitem{ref2}
 Wang Y and Voit J 1996
 An exactly solvable Kondo problem for interacting
 one-dimensional fermions
 \PRL {\bf 77} 4934--37;
 1997 \PRL {\bf 78} 3799 [Erratum]
\item[]
 Wang Y, Dai J, Hu Z-N and Pu F-C 1997
 Exact results for a Kondo problem in a one-dimensional $t$-$J$ model
 \PRL {\bf 79} 1901--4
\item[]
 Hu Z-N, Pu F-C and Wang Y 1998
 Integrabilities of the $t$-$J$ model with impurities
 \JPA {\bf 31} 5241--62

\bibitem{ref2w}
 Wang Y 1997
 Exact solution of the open Heisenberg chain with two impurities
 \PR B {\bf 56} 14045--9
\item[] Hu Z-N and Pu F-C 1998
 Two magnetic impurities in a spin chain
 \PR B 58  R2925--8
\item[]
 Chen S, Wang Y and Pu F-C 1998
 The open XXZ chain with boundary impurities
 \JPA {\bf 31} 4619--31
\bibitem{bi}
 Zhou H-Q, Ge X-Y, Links J and Gould M D 1998
 Graded reflection equation algebras and integrable Kondo impurities in the
 one-dimensional $t$-$J$ model
 {\it Nucl. Phys. B} {\bf 546} 779--91
\item[]
 Zhou H-Q, Ge X-Y and Gould M D 1999
 Integrable Kondo impurities in the one-dimensional supersymmetric $U$
 model of strongly correlated electrons
\JPA {\bf 32} L137-L142 {\it Preprint} cond-mat/9811049
\item[]
Fan H, Wadati M and Yue R 1999 Boundary Kondo impurities in the generalized
supersymmetric $t$-$J$ model
{\it Preprint} cond-mat/9906409

\bibitem{gradQISM}
Kulish P P and Sklyanin E K 1982 Solutions of Yang-Baxter equation 
J. Sov. Math. {\bf 19} 1596--620

\bibitem{ref11}
 G\"ohmann F and Murakami S 1998
 Fermionic representations of integrable lattice systems
 \JPA {\bf 31} 7729--49

\bibitem{ref6}
 Fedyanin V K and Yushankhay V 1978
 Soliton excitations in the model of a
 large-radius polaron
 {\it Theor. Math. Phys.} {\bf 35} 434--8
 [{\it Teor. Mat. Fiz.} {\bf 35} 240--6]
\item[]
 Makhankov V G and Fedyanin V K 1984
 Non-linear effects in quasi-one-dimensional models  of condensed
 matter theory  Phys. Rep. {\bf 104} 1--86
\item[]
 Pu F-C and Zhao B H 1986
 Exact solution of a polaron model in one dimension
 \PL A {\bf 118} 77--81

\bibitem{ref7}
 Zhou H-Q, Jiang L-J and Tang J-G 1990
 Some remarks on the Lax pairs for a one-dimensional small-polaron model
 and the one-dimensional Hubbard model
 \JPA {\bf 23} 213--23

\bibitem{ref10}
 Guan X-W, Grimm U and R{\"o}mer R A 1998
 Lax pair formulation for a small-polaron chain with integrable boundaries
 {\it Ann.\ Phys.\ (Leipzig)} {\bf 7} 518--22

\bibitem{ref4}
 Kulish P P 1996
 Yang-Baxter equation and reflection equations in integrable models
 {\it Low-Dimensional Models in Statistical Physics and Quantum Field
Theory}
 ed H Grosse and L Pittner (Berlin: Springer), pp.~125--44
\item[]
 Shiroishi M and Wadati M 1997
Integrable boundary conditions for the one-dimensional Hubbard model
\JPSJ {\bf 66} 2288--301
\item[]
 Bracken A J, Ge X-Y, Zhang Y-Z and Zhou H-Q 1998
 Integrable open-boundary conditions for the $q$-deformed
 supersymmetric $U$ model of strongly correlated electrons
 {\it Nucl. Phys.} B {\bf 516} 588--602

\bibitem{gfy}
 Guan X-W, Fan H and Yang S-D 1999
 Exact solution for small-polaron model with boundaries
\PL A {\bf 251} 79--85

\bibitem{reg}
 Frahm H and Slavnov N A 1999
 New solutions to the reflection equation and the projecting method
\JPA {\bf 32} 1547--55

\bibitem{KS} Kulish P P and Sklyanin E K 1982 
{\it Quantum spectral transform method, recent developments},
Lect. Notes Phys. {\bf 151} 61--119,
Berlin-New York: Springer
\bibitem{ref12}
 Alcaraz F C, Baake M, Grimm U and Rittenberg V 1987
 Operator content of the XXZ chain
 \JPA {\bf 20} L117--20
\item[]
 Hamer C J, Quispel G R W and Batchelor M T 1987
 Conformal anomaly and surface energy for Potts and Ashkin-Teller
 quantum chains
 \JPA {\bf 20} 5677--93
\item[]
 Alcaraz F C, Baake M, Grimm U and Rittenberg V 1989
 The modified XXZ Heisenberg chain, conformal invariance
 and the surface exponents of $c<1$ systems
 \JPA {\bf 22} L5--11
\item[]
 Woynarovich F, Eckle H-P and Truong T T 1989
 Non-analytic finite-size corrections in the one-dimensional Bose gas
 and Heisenberg chain
 \JPA {\bf 22} 4027--43
\item[]
 Grimm U and Sch\"{u}tz G 1993
 The spin-1/2 XXZ Heisenberg chain,
 the quantum algebra $U_{q}[sl(2)]$,
 and duality transformations for minimal models
 {\it J.\ Stat.\ Phys.} {\bf 71} 921--64

\bibitem{ref13}
 Eckle H-P and Hamer C J 1991
 Non-analytic finite-size corrections for the Heisenberg chain in a
 magnetic field with free and twisted boundary conditions
 \JPA {\bf 24} 191--202
\item[]
 Skorik S and Saleur H 1995
 Boundary bound states and boundary bootstrap in the sine-Gordon model
 with Dirichlet boundary conditions
 \JPA {\bf 28} 6605--22
\item[]
 Asakawa H and Suzuki M 1996
 Finite-size corrections in the XXZ model and the Hubbard model with
 boundary fields
 \JPA {\bf 29} 225--45
\item[]
 Kapustin A and Skorik S 1996
 Surface excitations and surface energy of the antiferromagnetic XXZ
 chain by the Bethe ansatz approach
 \JPA {\bf 29} 1629--38

\bibitem{KraM93} 
Kramer B and MacKinnon A 1993
Localization: theory and experiment
{\it Rep. Prog. Phys.} {\bf 56} 1469--564


\bibitem{Hec87}
 Hecht E 1987 
{\it Optics} (Addison-Wesley, Reading, Massachusetts, 2nd edition)

\endbib
\end{document}